\documentstyle[preprint,prd,aps,eqsecnum]{revtex}

\begin{document}
\preprint{DART-HEP-93/02\hspace{-26.6mm}\raisebox{-2.4ex}{ April 1993}}




\title{On the Evaluation of Thermal Corrections to False Vacuum Decay Rates}



\author{Marcelo Gleiser$^{1)}$, Gil C. Marques\thanks{On leave from Inst.
de F\'{\i}sica, Univ.
de S\~ao Paulo, C.P. 20516, S\~ao Paulo, SP 01498, Brazil.}
$^{2)}$,
and Rudnei O. Ramos$^{1)}$}

\address{{\it $^{1)}$ Department of Physics and Astronomy, Dartmouth College}\\
Hanover, NH 03755, USA}

\address{{\it $^{2)}$ Department of Physics, Texas A $\&$ M University }\\
{\it College Station, TX 77843, USA}}


\maketitle

\begin{abstract}
\baselineskip 12pt

We examine the computation of the nucleation barrier used in the
expression for false vacuum decay rates in finite temperature field theory.
By a detailed analysis of the determinantal prefactor, we show that the
correct bounce solution used in the computation of the nucleation barrier
should
not include loop corrections coming from the scalar field undergoing decay.
Temperature corrections to the bounce appear from loop contributions from other
fields coupled to the scalar field. We compute the nucleation barrier
for a model of scalar
fields coupled to fermions, and compare our results to the expression commonly
used in the literature. We find that, for large enough self-couplings,
the inclusion of scalar loops in the
expression of the nucleation barrier leads to an underestimate of the decay
rate in the neighborhood of the critical temperature.

\vspace{0.21in}

\noindent
PACS number(s): 98.80.Cq, 64.60.Qb .\\

\noindent
e-mail: gleiser@peterpan.dartmouth.edu; marques@phys.tamu.edu;
\break rudnei@northstar.dartmouth.edu .

\end{abstract}

\def\vp{\varphi}
\def\p{\phi}

\newpage

\section{\bf Introduction}


For the past decade or so, the study of first-order phase
transitions in cosmology has been the focus of much
interest due to their possible relevance to the
physics of the early Universe. Some well-known examples are
inflationary models \cite{kolb}, the quark-hadron transition
\cite{qcd} and, more
recently, the generation of the cosmological baryon asymmetry in
the electroweak phase transition \cite{weak}.

In a first-order phase transition, the initial metastable phase decays to
the stable phase by the nucleation of bubbles larger than a critical size.
This decay may be triggered by either quantum or thermal fluctuations,
depending on how the ambient temperature compares to the nucleation barrier
\cite{gunton}.
Within a cosmological context, the cooling  is provided by the expansion
of the Universe; the long-wavelength modes of the order parameter responsible
for the symmetry breaking transition are coupled to the ``environment'', which
is assumed to be in local thermal equilibrium at some temperature $T$.
(Here, we are mainly concerned with ``late'' transitions, for which the
typical fluctuation time-scales are much shorter than the expansion rate.)

Of great relevance to the understanding of the evolution of the
phase transition is the determination of the bubble nucleation rate
per unit volume. This is a well-known problem in classical statistical
mechanics, with an long history \cite{gunton}. Phenomenological
field-theoretic
treatments were developed by Cahn and Hilliard \cite{cahn}, and
by Langer \cite{langer67} in the context of a time-dependent
coarse-grained Ginzburg-Landau model. Classical homogeneous
nucleation theory within a
field theoretic context has been recently shown by numerical experiments
to successfully predict the
nucleation barrier \cite{alford}.
In the case of zero-temperature quantum field
theory, the study of metastable vacuum decay was initiated with the work of
Voloshin, Kobzarev, and Okun \cite{volosh}, and was put onto firm
theoretical ground by Coleman and Callan in the late seventies \cite{coleman}.
Finite temperature corrections to the vacuum decay
rate were first considered by
Linde \cite{linde}, who argued that
temperature corrections to the nucleation rate are obtained recalling that
finite temperature field theory (at sufficiently high temperatures) in
$d+1$-dimensions
is equivalent to $d$-dimensional euclidean quantum field theory
with $\hbar$ substituted by $T$. Thus, in $d+1$ dimensions, the
nucleation rate is proportional to ${\rm exp} [-S^d_E(\vp_b)/T]$, where
$S^d_E(\vp_b)$ is the $d$-dimensional euclidean action evaluated at
its extremum
(specifically, a saddle point),
the critical
bubble, or bounce,  $\vp_b(r)$. The usual expression for the nucleation
rate per unit
volume used in the literature is \cite{linde}

\begin{equation}
\Gamma = T \left( \frac{ S_E^3 (\vp_b(r,T),T)}{2 \pi T}
\right)^{\frac{3}{2
}}
\left\{ \frac{ \det [ - \nabla^{2} + V_{\rm {eff}}''(\varphi_{f},T)]}{
\det' [ - \nabla^{2} + V_{\rm {eff}}''(\varphi_b(r,T),T)]} \right\}^
{\frac{1}{2}}
{\rm exp}\left [- {{S_E^3 (\varphi_b(r,T),T)}\over T}\right ] \: ,
\label{badrate}
\end{equation}
\noindent
where $\vp_f$ is the value of the field $\phi$ at the metastable minimum, and
the prime in the determinant in the denominator is a reminder that one
should omit the zero and negative eigenvalues, associated with the
translation symmetry of the bubble and with its instability (being a saddle
point configuration), respectively.
$V_{\rm eff}(\phi,T)$ is the one-loop approximation to the
finite temperature effective potential,
and $V_{\rm eff}''(\varphi,T)= \frac{\partial^{2} V_{\rm eff}(\phi,T)}{\partial
\phi^{2}}
|_{\phi =
\varphi}$.

There are three important points here. The first is that in order to
estimate the determinantal prefactor (the ``equilibrium'' part of the
prefactor; there is a dynamic factor which can not be obtained by
using equilibrium arguments) one usually
proceeds by invoking dimensional arguments
to approximate it by a term of order $T_C^4$
($T_C$ is the critical temperature). How
good is this approximation? Clearly, in most cases it is impossible to
evaluate the determinants exactly. But can one obtain a better approximation
than the simple use of dimensional arguments?
The second, and most important, point is that the critical bubble configuration
$\vp_b(r,T)$ used to evaluate the nucleation barrier, denoted above by
$S_E^3(\vp_b(r,
T))$, was obtained from an effective potential  which includes corrections
coming from scalar loops. Hence the temperature dependence in $\vp(r,T)$.
We will show here that this procedure
is not in general
justified and is only a good approximation if the corrections from scalar
loops are negligible. Finally, in the expression for the
temperature corrected barrier in
Eq. (\ref{badrate}),
$S^3_E(\vp_b(r,T),T)$, one uses the {\it temperature corrected}
effective potential, $ V_{\rm eff}(\vp,T)$
as opposed to the tree level potential. Thus, it is claimed that
$S^3_E(\vp_b(r,T),T)$ is equivalent to the
free energy of the temperature dependent bounce, given by
\begin{equation}
S^{3}_E (\varphi_b(r,T),T) = \int d^{3} x \left[ \frac{1}{2} \left( \nabla
\varphi_b(r,T)\right)^{2} + V_{\rm eff}( \varphi_b(r,T),T) \right] \: .
\label{badact}
\end{equation}
As far as we know, apart from the work of Affleck in the context of finite
temperature quantum mechanics \cite{affleck}, this point has never been
properly addressed in the literature. How did the temperature
corrected potential appear in the exponent? Is the exponentiation of the
massless modes sufficient?
In fact, most of the work done
on cosmological phase transitions in which temperature effects are
important (including some by the present authors)
simply invokes Linde's results.
Given the many applications of finite temperature vacuum decay in cosmology, we
feel that this important question should not be left unscrutinized. This
concern has also been expressed
in recent works by Csernai and Kapusta \cite{csernai},
and by Buchm$\ddot{\rm u}$ller, Helbig, and Walliser \cite{buch}. Both works
attempted to improve on Linde's results, by generalizations of Langer's
work. Csernai and Kapusta obtained an expression for the dynamical prefactor
by using a relativistic hydrodynamic approach, while Buchm$\ddot{\rm u}$ller
{\it et al.} obtained an approximate expression for the decay rate in
scalar electrodynamics (and more recently, with Z. Fodor,
in the standard electroweak model)
by integrating out the electromagnetic degrees of
freedom from the partition function.
However, a more detailed analysis of the nucleation barrier and how it
compares to the usual result is still lacking.

In this paper we address the three points raised above.
We will be mostly interested in the
regime in which thermal fluctuations are much larger than quantum
fluctuations. This way we avoid the question of how to match
continuously the two regimes,
although we believe this
to be a very important question \cite{leggett}. (See also Refs. \cite{affleck}
and \cite{igor}.)
The hope is that in most situations of
interest the
transition will be dominated by one or the other regime.
By a saddle-point evaluation of the
partition function in the case of a self-interacting scalar field,
it is possible to show that the temperature corrections
to the nucleation barrier can be interpreted as entropic contributions due
to stable vibrational modes on the {\it tree-level} bounce configuration,
$\vp_b(r)$.
In other words, the
first corrections to the energy of the critical bubble configuration come from
temperature induced stable fluctuations on the bubble, which
will modify its volume and surface energies. We will show that
these corrections are given by the temperature corrected effective
action evaluated at the tree-level bounce $\vp_b(r)$.
In expression (\ref{badrate}),
the bounce is obtained from the effective potential which includes
scalar loops.
The difference between the two nucleation barriers
will be important whenever scalar loops are not negligible. We
will show that they become particularly important within the so-called
thin-wall limit, that is, in the vicinity of the critical temperature for
the transition. This is perfectly consistent with the fact that large
entropic corrections are expected near the critical temperature.
We will obtain this result
by a perturbative evaluation of the
determinantal prefactor. In principle, the determinantal
prefactor
can be evaluated in two ways. Clearly, the computation can be done directly
if we know the eigenvalues related to a
given bubble configuration.
This method is not very useful in practice, since we in general
do not know the eigenvalues. (Unless, of course, we obtain them numerically.)
Writing down explicitly the eingenvalue equations,
and using a thin-wall approximation
to the bubble configuration, we show how the temperature corrections
to the nucleation barrier originate from fluctuations about the critical bubble
configuration. Even though the thin-wall approximation is not very useful
in realistic situations, there is no reason to believe that
thicker wall bubbles will
behave any differently. (Unless the transition becomes too weak, in which
case nucleation of critical bubbles may not be the relevant mechanism for
the transition \cite{gkw}.)
The second approach we use to evaluate the prefactor relies on a perturbative
expansion of the determinants. Within first-order, it is again possible to show
how the prefactor accounts for the temperature corrections to the nucleation
barrier.

The paper is organized as follows.
In Section 2 we briefly review
Langer's formalism for obtaining nucleation rates, adapted to field
theory at finite temperatures. That is, we obtain the partition function
for the metastable phase plus a nucleating fluctuation
by a saddle-point evaluation of the functional integral.
In Section 3 we show how the determinantal prefactor can account for the
finite temperature corrections to the nucleation barrier. For simplicity,
the calculation
is performed in the context of the thin-wall approximation for the bubble
profile, although in principle one could obtain results for any configuration.
In Sections 2 and 3, for the sake of clarity, the discussion is
somewhat oversimplified. We {\it assume} that
the system is initially in  a metastable state and study only the scalar
degrees of freedom in the problem. This situation is not unrealistic, as it
can be reproduced in numerical simulations of vacuum decay \cite{alford}.
In Section 4 we study a model of a scalar field coupled to fermions. This
example is particularly interesting as it illustrates how a thermal state
evolves into a metastable
state due to radiative corrections, very much like in the standard electroweak
model. (Recall that in a cosmological context
the cooling is provided by the expansion of the Universe.)
A related problem has
been recently studied by E. Weinberg, in the context of massless scalar
models for which symmetry breaking occurs due to radiative corrections.
In the regime dominated by quantum fluctuations, Weinberg showed how
radiative corrections
induce a metastable state and how it is possible to
evaluate its decay rate\cite{weinb}.
We show how to obtain an
effective
partition function for the scalar field by integrating out the fermionic
modes, and proceed to obtain the decay rate. We then compare our results
for the nucleation barrier to the results obtained using Eq. (\ref{badrate}).
Conclusions are presented in Section 5 and two Appendices are included to
clarify a few technical points.

\section{\bf FINITE TEMPERATURE DECAY RATE: GENERAL FORMALISM}

Consider a scalar field model with four-dimensional euclidean action

\begin{equation}
S_{\rm E}=\int d^4x_E~ {\cal L}_E  \:,
\label{Seucl}
\end{equation}
\noindent
where ${\cal L}_E$ is the euclidean lagrangian density given by
\begin{equation}
{\cal L}_E = \frac{1}{2} (\partial_{\mu} \phi)^{2} + V(\phi)
  \: ,
\label{LphiH}
\end{equation}

\noindent
and the potential in (\ref{LphiH}) has a metastable
minimum at $\phi=\vp_f$ and
a stable minimum at $\phi=\vp_t$, as shown in Fig. 1. [We will only consider
potentials with two minima here.]

Let us {\it assume} that the system is prepared initially in the metastable
phase, without worrying for the moment about how this is done. (See Section 4.)
The metastable phase will
decay into the stable phase by the nucleation of bubbles larger than a
critical size. (For a review see, e.g., \cite{gunton}.) As is well-known,
in order to study the decay of the false vacuum at finite temperature we
impose the periodic boundary condition (anti-periodic for fermions)
$\phi(0,{\bf x})=\phi(\beta\hbar, {\bf x})$, so that the euclidean action
becomes \cite{kapusta}

\begin{equation}
S_E[\phi]=\int_0^{\beta}d\tau\int d^3x\left [{1\over 2}
\left ({{\partial \phi}\over
{\partial \tau}}\right )^2+{1\over 2} \left (\nabla\phi \right )^2 +V(\phi)
\right ] \:.
\label{SeuclT}
\end{equation}

\noindent
The partition function of the system is given by a functional integral over
all possible field configurations weighted by their euclidean action,

\begin{equation}
Z = \int D \phi e^{- S_{E}(\phi)} \:.
\label{ZphiH}
\end{equation}

Following Langer \cite{langer67}, we describe the nucleation
of bubbles of the stable phase inside the metastable phase
under the assumption that a dilute gas approximation for these
droplets is valid. Unless the transition is weakly first-order, this should be
a very good approximation to describe the early stages of the transition,
when bubble collisions and other complicated kinetic effects can be neglected.
The critical
configuration is an extremum of the euclidean action,

\begin{equation}
\frac{\delta S_{E}(\phi)}{\delta \phi}|_{\phi = \varphi_{b}} = 0 \:,
\label{extrem}
\end{equation}

\noindent
being thus a solution of the equation of motion,

\begin{equation}
( \frac{\partial^{2}}{\partial \tau^{2}} + \nabla^{2} ) \phi =
V'(\phi) \:,
\label{eulereq}
\end{equation}

\noindent
with boundary conditions, ${\rm lim}_{\tau\rightarrow \pm \infty}\phi(\tau,
\vec {x})=\vp_f$, and ${\rm lim}_{|\vec {x}|\rightarrow \infty}=\vp_f$.
Coleman, Glaser, and Martin \cite{cgm}, have shown that
the configuration with minimum
energy, {\it i.e.}, the minimum of all the maxima, will have $O(4)$-symmetry.
As argued by Linde \cite{linde}, for sufficiently high temperatures the
problem becomes effectively three-dimensional, and the saddle point will be
given by the $O(3)$-symmetric, or static, solution of

\begin{equation}
\frac{d^{2} \phi}{d r^{2}} + \frac{2}{r} \frac{d \phi}{d r} = V'(\phi)
\: ,
\label{eulersph}
\end{equation}

\noindent
with boundary conditions, $\lim_{r \to +\infty} \phi = \varphi_{f}$ and
$\frac{d \phi}{d r}|_{r=0} = 0$. Note that the potential in (\ref{eulersph})
is the {\it tree-level} potential.  For future reference we note that
when the false vacuum energy density
[$\Delta V\equiv V(\vp_f)-V(\vp_t)$]
is much smaller than the barrier height [$V_h$, see Fig. 1],
the bubble radius $R$ is much larger than the wall thickness $\Delta R \sim
m^{-1}$, where $m$ is a typical mass scale in the problem.
In this case, the solution to Eq. (\ref{eulersph}) can be estimated by the
so-called
thin-wall approximation and is given by

\begin{equation}
\varphi_{b}(r) = \left\{
\begin{array}{ll}
\varphi_{t},  & 0 < r < R- \Delta R \\
\varphi_{wall}(r),  & R - \Delta R < r < R + \Delta R \\
\varphi_{f},     & r >  R + \Delta R
\end{array}
\right.
\label{bubble}
\end{equation}

\noindent
which describes a bubble of radius $R$ of the stable phase $\varphi_{t}$
embedded in the metastable phase $\varphi_{f}$. $\varphi_{wall}(r)$ describes
the bubble wall separating the two phases.

The main advantage of Langer's approach is that in a dilute gas of bubbles
one can infer the thermodynamics of the system from the knowledge of the
partition function of a single bubble. We can write the
partition function for the system with a bubble of the stable vacuum inside
the metastable vacuum as

\begin{equation}
Z = Z(\varphi_{f}) + Z(\varphi_{b}) \: ,
\label{Z1bubb}
\end{equation}

\noindent
where $Z(\varphi_{f})$ and $Z(\varphi_{b})$ are the partition functions of
the system for the vacuum field configuration $\varphi_{f}$ and for
the bubble field configuration $\varphi_{b}$, respectively \cite{ramos}.
The generalization of (\ref{Z1bubb}) for several bubbles is given by

\begin{eqnarray}
Z & \simeq & Z(\varphi_{f}) +Z(\varphi_{f}) \left[ \frac{Z(\varphi_{b})}
{Z(\varphi_{f})} \right] + Z( \varphi_{f}) \frac{1}{2 !} \left[
\frac{Z(\varphi_{b})}{Z(\varphi_{f})} \right]^{2} + \ldots
\nonumber \\
& \simeq & Z(\varphi_{f}) \exp \left[ \frac{Z(\varphi_{b})}{Z(\varphi_{f})}
\right]
\: .
\label{Zmany}
\end{eqnarray}

\noindent
The proof of (\ref{Zmany}) can be found in the work
of Arnold and McLerran \cite{mclerran}, who studied the properties
of a dilute gas of sphalerons. They expressed the multiple-sphaleron
configurations
as the superposition of many single sphalerons, with partition
function approximated as above.

The partition functions in (\ref{Zmany})
can be evaluated by the saddle-point method, expanding the lagrangian
field in (\ref{ZphiH}) as
$\phi(\vec{x} ,\tau) \rightarrow \varphi_{b} (\vec{x}) +\eta (\vec{x} ,\tau)$
for $Z(\varphi_{b})$ and  $\phi(\vec{x} ,\tau) \rightarrow  \varphi_{f} +
\zeta(\vec{x} ,\tau)$  for
$Z(\varphi_{f})$. $\eta(\vec{x} ,\tau)$ and $\zeta(\vec{x} ,\tau)$ are small
perturbations around the classical field configurations $\varphi_{b} (\vec{x})$
and  $\varphi_{f} $, respectively. These perturbations around
each configuration bring the temperature corrections to the nucleation barrier
into the problem, as we shall see.
Up to 1--loop order one keeps the quadratic
terms in the fluctuations $\eta(\vec{x} ,\tau)$ and $\zeta(\vec{x} ,\tau)$
in the
expansion of the scalar field in the partition function (\ref{ZphiH}).
This way one can write the following expressions for
$Z(\varphi_{b})$ and  $Z(\varphi_{f})$, respectively,

\begin{equation}
Z(\varphi_{b}) \stackrel{1-loop \: order}{\simeq} e^{-S_{E}(\varphi_{b})}
\int D \eta \exp \left\{ -\int_{0}^{\beta} d \tau \int d^{3} x \frac{1}{2}
\eta \left[ -\Box_E + V''(\varphi_{b}) \right] \eta \right\}
\label{Zphib}
\end{equation}

\noindent
and

\begin{equation}
Z(\varphi_{f}) \stackrel{1-loop \: order}{\simeq} e^{-S_{E}(\varphi_{f})}
\int D \zeta \exp \left\{ -\int_{0}^{\beta} d \tau \int d^{3} x \frac{1}{2}
\zeta \left[ -\Box_{E} + V''(\varphi_{f}) \right] \zeta \right\} \: ,
\label{Zphif}
\end{equation}

\noindent
where
$V''(\varphi) = \frac{d^{2} V(\phi)}{d \phi^{2}}|_{\phi = \varphi}$
and $\Box_{E} = \frac{\partial^{2}}{\partial \tau^{2}} +
\vec{\nabla}^{2}$ .

Performing the functional Gaussian integrals in (\ref{Zphib}) and (\ref{Zphif})
one gets the
following expression for the ratio between the partition functions,
$\frac{Z(\varphi_{b})}{Z(\varphi_{f})}$, appearing in (\ref{Zmany}):

\begin{equation}
\frac{Z(\varphi_{b})}{Z(\varphi_{f})} \stackrel{1-loop \: order}{\simeq}
\left[ \frac{\det ( -\Box_{E} + V''(\varphi_{b}))_{\beta}}
{ \det ( -\Box_{E} + V''(\varphi_{f}))_{\beta}} \right]^{-\frac{1}{2}}
e^{-\Delta S} \: ,
\label{ratiodet}
\end{equation}

\noindent
where $[ \det (M)_{\beta}]^{- \frac{1}{2}} \equiv \int D \eta \exp \left\{
- \int_{0}^{\beta} d \tau \int d^{3} x \frac{1}{2} \eta [M] \eta \right\}$
and  $\Delta S = S_{E}(\varphi_{b}) - S_{E}(\varphi_{f})$
is the difference
between the euclidean actions for the field configurations
$\varphi_{b}$ and $\varphi_{f}$. Note that $S_{E}(\vp)$, and hence $\Delta S$,
{\it does not include any temperature corrections.}

{}From (\ref{Zmany}) and (\ref{ratiodet}),
the free energy of the system, ${\cal F} = - \beta^{-1} \ln Z$
can be written as

\begin{equation}
{\cal F} = - T \left[ \frac{ \det ( -\Box_{E} + V''(\varphi_{b}))_{\beta}}
{ \det ( -\Box_{E} + V''(\varphi_{f}))_{\beta}} \right]^{-\frac{1}{2}}
e^{-\Delta S} \: .
\label{energy}
\end{equation}

As is well-known, the determinantal prefactor evaluated
for the bounce configuration
has one negative eigenvalue,
signalling the presence of a metastable state, and also three
zero eigenvalues related with the translational invariance of the bubble
in three-dimensional space. Because of the negative eigenvalue, the free
energy ${\cal F}$ is imaginary. As shown by Langer \cite{langer67}
(see also Ref. \cite{affleck}),
the decay rate is proportional to the imaginary part of
${\cal F}$

\begin{equation}
{\cal R}= \frac{| E_- |}{\pi T} {\rm Im} {\cal F} \: ,
\label{gamma}
\end{equation}

\noindent
where $| E_-|$ is the single negative eigenvalue. In general it depends on
non-equilibrium aspects of the dynamics, such as the coupling strength to the
thermal bath.

\section{\bf Evaluation of the Determinants}

In this Section we compute the ratio of the determinants appearing in
the decay rate, and show how it provides
a finite temperature correction to the nucleation barrier.

First recall that for static field configurations $\Delta S$ is given by

\begin{equation}
\Delta S = \beta \int d^{3} x \left[ {\cal L}_{E}(\varphi_{b}) -
{\cal L}_{E}(\varphi_{f}) \right] = \frac{\Delta E}{T} \: ,
\label{deltaS}
\end{equation}

\noindent
where $\Delta E$ is simply the nucleation {\it energy} barrier, that is, the
energy of a critical nucleation within the metastable phase.
For example, in the thin-wall approximation of Eq. (\ref{bubble}),
$\Delta E$ is

\begin{equation}
\Delta E = - \frac{4 \pi R^{3}}{3} \Delta V + 4 \pi R^{2} \sigma_{0} \: ,
\label{deltaE}
\end{equation}

\noindent
where
$\sigma_{0}$ is the tree-level surface tension of
the bubble wall ({\it i.e.} with no corrections due to fluctuations
around the bubble
wall field configuration $\varphi_{wall}$)

\begin{equation}
\sigma_{0} \simeq \int_{-\Delta R}^{+ \Delta R} d r \left[
{\cal L}_{E}(\varphi_{wall}) - {\cal L}_{E}(\varphi_{f})
\right] \: .
\label{sigma0}
\end{equation}

Using (\ref{deltaS}) in Eqs. (\ref{energy}) and (\ref{gamma}) we can write
the decay rate as

\begin{equation}
{\cal R}= -{{|E_-|}\over {\pi}}{\rm Im}
\left [ \frac{ \det ( -\Box_{E} + V''(\varphi_{b}))_{\beta}}
{ \det ( -\Box_{E} + V''(\varphi_{f}))_{\beta}} \right ]^{-\frac{1}{2}}
{\rm exp} \left (-{{\Delta E }\over T}\right ) \:.
\label{gamma2}
\end{equation}

The determinantal prefactor in (\ref{gamma2}) will provide the temperature
corrections
to the nucleation barrier. This should come as no surprise, given that
the determinant is obtained from integrating over thermally induced
fluctuations about the tree-level
bubble configuration.
(Recall that we are only considering the regime in
which thermal fluctuations are much
larger than quantum fluctuations.)
Once the negative and zero eigenvalues are taken care of,
the positive eigenvalues are easily associated with entropic contributions to
the activation energy due to stable deformations of the bubble's shape, as in
classical nucleation theory.

We now proceed to show how to incorporate temperature corrections to
the nucleation barrier. This can be done without an explicit evaluation of
the eigenvalues of the operators in the determinants. As we show next, all
that we need is to
separate consistently the positive eigenvalues from the negative and zero
eigenvalues, and then show how the former can be exponentiated.
In principle, the computation of the determinantal prefactor in (\ref{energy})
can be performed by two different methods.
The first involves obtaining directly (analytically, or
more realistically, numerically) the eigenvalues
for the determinants in Eq. (\ref{energy}).
The second method consists in developing
a consistent perturbative expansion for the ratio of the determinants. We now
examine both these possibilities.

\subsection{\bf Evaluating the Determinantal Prefactor: Eigenvalue Equations}

Consider the eigenvalue equations for the differential operators that appear
in the determinantal prefactor,
\begin{equation}
\left[ - \Box_E + V''(\varphi_{f}) \right] \Phi_{f}(i) =
\varepsilon_{f}^{2}(i) \Phi_{f}(i)
\label{valuephif}
\end{equation}

\noindent
and

\begin{equation}
\left[ - \Box_E + V''(\varphi_{b}) \right] \Phi_{b}(i) =
\varepsilon_{b}^{2}(i) \Phi_{b}(i) \: .
\label{valuephib}
\end{equation}

\noindent
In momentum space one writes, $\varepsilon^{2} = \omega_{n}^{2} + E^{2}$,
where $\omega_{n} = \frac{2 \pi n}{\beta} \:,\:\:n = 0,\pm 1,\pm 2, \ldots $,
for bosons (for fermion fields $\omega_{n} = \frac{(2 n + 1) \pi}{\beta}$).
{}From (\ref{valuephif}) and (\ref{valuephib})
one can write the determinant ratio in (\ref{energy}) as

\begin{eqnarray}
\left[ \frac{ \det ( -\Box_{E} + V''(\varphi_{f}))_{\beta}}
{ \det ( -\Box_{E} + V''(\varphi_{b}))_{\beta}} \right]^{\frac{1}{2}}
&=& \exp \left\{ \frac{1}{2} \ln \left[ \frac{ \det ( -\Box_{E} +
V''(\varphi_{f}))_{\beta}}
{ \det ( -\Box_{E} + V''(\varphi_{b}))_{\beta}} \right] \right\} =
\nonumber \\
&=& \exp \left\{ \frac{1}{2} \ln \left[ \frac{ \prod_{n=- \infty}^{+ \infty}
\prod_{i} \left( \omega_{n}^{2} + E_{f}^{2}(i) \right) }{
\prod_{n= - \infty}^{+ \infty} \prod_{j} \left( \omega_{n}^{2} + E_{b}^{2}(j)
\right) } \right] \right\} \: .
\label{ratio}
\end{eqnarray}

Using the identity,

\begin{equation}
\prod_{n = 1}^{+ \infty} \left( 1 + \frac{z^{2}}{n^{2}} \right) =
\frac{\sinh (\pi z)}{ \pi z}
\label{idenpi}
\end{equation}

\noindent
and taking into account that we have in the denominator of
(\ref{ratio}) one negative and
three zero eigenvalues, one obtains (for details see Appendix A)

\begin{eqnarray}
\left[ \frac{ \det ( -\Box_{E} + V''(\varphi_{f}))_{\beta}}
{ \det ( -\Box_{E} + V''(\varphi_{b}))_{\beta}} \right]^{\frac{1}{2}}
&=& {\cal V}\frac{T^{4}}{i |E_{-}| } \frac{\beta \frac{|E_{-}|}{2}}{
\sin \left( \beta \frac{|E_{-}|}{2} \right)} \left[ \frac{\Delta E}{2 \pi T}
\right]^{\frac{3}{2}}  \nonumber \\
& \times & \exp  \left\{ \sum_{i} \left[ \frac{\beta}{2} E_{f}(i) +
\ln \left( 1 - e^{- \beta E_{f}(i)} \right) \right]  \right. \nonumber \\
&-& \left.
\sum_{j} \; ' \left[ \frac{\beta}{2} E_{b}(j) + \ln \left( 1 -
e^{- \beta E_{b}(j)} \right) \right] \right\} \: .
\label{eigenratio}
\end{eqnarray}

\noindent
The factor ${\cal V}\left[ \frac{\Delta E}{2 \pi T}\right]^{\frac{3}{2}} $
in the right hand side of Eq. (\ref{eigenratio}) comes from the contribution of
the zero eigenvalues, which can be handled as in
ref. \cite{coleman}, through the use of collective coordinates corresponding
to the position of the bubble. ${\cal V}$ is the volume of three space,
while the prime in $\sum_{j}$ is a reminder that
we have excluded the negative and the three zero eigenvalues from
the sum.
Thus, the argument of the exponential incorporates
only the contributions from the stable vibrational modes on the bubble, very
much like in Langer's result \cite{langer67}.

Substituting Eq. (\ref{eigenratio}) in (\ref{gamma2})
one obtains the following expression for
the nucleation rate per unit volume, $\Gamma\equiv {\cal R}/{\cal V}$,
as defined in Eq. (\ref{gamma}):

\begin{equation}
\Gamma = {\cal A} T^{4}
\exp \left[ - \frac{ \Delta F(T)}{T} \right] \: ,
\label{endgamma}
\end{equation}

\noindent
where we have denoted by ${\cal A}$ the dimensionless factor

\begin{equation}
{\cal A} = \frac{1}{\pi} \frac{ \frac{|E_{-}|}{2 T}}{ \sin \left( \frac{
|E_{-}|}{2 T} \right)} \left[ \frac{\Delta E}{2 \pi T} \right]^{\frac{3}{2}}
\:.
\label{calA}
\end{equation}

\noindent
Note that the zero-mode
contribution in the prefactor depends on the energy barrier of the critical
nucleation, and not
on its free energy as in Eq. (\ref{badrate}).
In Eq. (\ref{endgamma}) we have incorporated
the exponential contribution from the prefactor into the
definition of the temperature corrected nucleation barrier, which we call
$\Delta F(T)$. Within the thin-wall approximation we can write

\begin{equation}
\Delta F(T) = - \frac{4 \pi R^{3}}{3} \Delta V_{\rm eff}(T) + 4 \pi R^{2}
\sigma (T) \:,
\label{deltaF}
\end{equation}

\noindent
where

\begin{eqnarray}
\Delta V_{\rm eff}(T) &=& V(\varphi_{f}) - V(\varphi_{t}) +
T \int \frac{d^{3} k}{(2 \pi)^{3}}
\ln \left[ 1 - e^{- \beta \sqrt{ \vec{k}^{2} + m^{2}(\varphi_{f})}} \right]
- \nonumber \\
&-& T \int \frac{d^{3} k}{(2 \pi)^{3}} \ln
\left[ 1 - e^{- \beta \sqrt{ \vec{k}^
{2} + m^{2}(\varphi_{t})}} \right]
\label{deltaU}
\end{eqnarray}

\noindent
and

\begin{eqnarray}
\sigma (T) &=& \sigma_{0} + \frac{T}{ 4 \pi R^{2}} \left\{ \sum_{j} \; '
\left[ \frac{ \beta}{2} E_{wall}(j) + \ln \left( 1 - e^{- \beta E_{wall}(j)}
\right) \right] - \right. \nonumber \\
&-& \left. \sum_{i} \left[ \frac{ \beta}{2} E_{f}(i) + \ln \left( 1 -
e^{- \beta
E_{f}(i)} \right) \right] \right\} \: .
\label{sigmaT}
\end{eqnarray}

\noindent
In Eq. (\ref{deltaU}) we have
substituted the discrete sums by integrals over momenta. [For
field configurations $\varphi_{f}$ and $\varphi_{t}$, we have the continuum
eigenvalues, $E_{f}^{2} = \vec{k}^{2} + m^{2}(\varphi_{f})$ and
$E_{t}^{2} = \vec{k}^{2} + m^{2}(\varphi_{t})$, respectively, with
$m^{2}(\varphi_{f}) = \frac{d^{2} V(\phi)}{d \phi^{2}} |_{\phi = \varphi_{f}}$
and
$m^{2}(\varphi_{t}) = \frac{d^{2} V(\phi)}{d \phi^{2}}
|_{\phi = \varphi_{t}}$.] We have omitted
the usual zero-temperature ultraviolet divergent
terms,
$\int d^{3} k \sqrt{ \vec{k}^{2} + m^{2}
(\varphi)}$,
from Eq. (\ref{deltaU})
since they can always be handled by the introduction of
the usual counterterms \cite{jackiw}.

In Eq. (\ref{sigmaT}) $E_{wall}(j)$ are the eigenvalues
related with the bubble wall
field configuration $\varphi_{wall}$. Thus, within the thin-wall
approximation, the problem is reduced to the
computation of these eigenvalues for a field configuration describing the
bubble wall, a non-trivial task.
It is possible, however, that
the temperature corrections to the surface density are negligible. For
example,
in the context of the QCD transition, the surface density is obtained from
lattice calculations, and is shown not to be very sensitive to temperature
\cite{csernai}.

It is easy to see from Eq. (\ref{deltaU}) that $\Delta V_{\rm eff}(T)$
is the usual 1-loop approximation to the finite
temperature false vacuum energy density \cite{jackiw}.
The second term in the right hand side of (\ref{sigmaT})
comes from the finite temperature 1-loop contribution to the surface
tension $\sigma_{0}$,
due to thermal fluctuations on the bubble wall. Thus, by exponentiating the
contribution from the stable modes, the activation energy for the critical
bubble becomes indeed an activation free energy. Note however, that
contrary to Eq. (\ref{badrate}), the free-energy functional is evaluated
for the tree-level bounce. In Section 4 we will compare the results obtained
with the two approaches.

\subsection{\bf Evaluating the Determinantal Prefactor: Perturbative Expansion}

A second approach to the computation of the determinantal prefactor
in (\ref{gamma2}) consists in developing a perturbative
expansion for it.
First we write the determinantal ratio as

\begin{eqnarray}
\left[ \frac{ \det ( -\Box_{E} + V''(\varphi_{f}))_{\beta}}
{ \det' ( -\Box_{E} + V''(\varphi_{b}))_{\beta}} \right]^{\frac{1}{2}}
&=& \exp \left\{ \frac{1}{2} {\rm Tr} \; \ln \left[-\Box_{E} +
V''(\varphi_{f}) \right]_{\beta} - \right. \nonumber \\
&-& \left. \frac{1}{2} {\rm Tr'} \; \ln \left[-\Box_{E} +
V''(\varphi_{b}) \right]_{\beta} \right\} \: ,
\label{fieldratio}
\end{eqnarray}

\noindent
where we have used in (\ref{fieldratio}) the identity
$ \ln \det \hat{M} = {\rm Tr} \; \ln
\hat{M}$ and the prime in both sides denote that the negative and the zero
modes have been omitted. (They are treated as in previous Section.)

We rewrite (\ref{fieldratio}) as

\begin{equation}
\left[ \frac{ \det ( -\Box_{E} + V''(\varphi_{f}))_{\beta}}
{ \det' ( -\Box_{E} + V''(\varphi_{b}))_{\beta}} \right]^{\frac{1}{2}}
= \exp \left\{ - \frac{1}{2} {\rm Tr} \; \ln \Bigl [ 1 + G_{\beta}(\varphi_{f})
\left[ V''(\varphi_{b}) - V''(\varphi_{f}) \right] \Bigr ] \right\} \: ,
\label{expratio}
\end{equation}

\noindent
where

\begin{equation}
G_{\beta}(\varphi_{f}) = \frac{1}{ - \Box_{E} + m^{2}(\varphi_{f}) }
\label{Gbetaphi}
\end{equation}

\noindent
is just the propagator for the scalar field
$\phi$, with
$m^{2}(\varphi_{f}) = V''(\varphi_{f})$.

Expanding the logarithm in (\ref{expratio})
in powers of \break $G_{\beta}(\varphi_{f})
\left[ V''(\varphi_{b}) - V''(\varphi_{f}) \right]$, we obtain the graphic
representation,

\def\ols{ \begin{picture}(77,22)(0,+8.5)
     \thicklines
  \multiput(4,11)(8,0){3}{\line(1,0){3}}
\put(36,11){\circle{25}}
\end{picture} }
\def\olr{ \begin{picture}(77,22)(0,+8.5)
     \thicklines
  \multiput(4,11)(8,0){3}{\line(1,0){3}}
\put(36,11){\circle{25}}
  \multiput(48,11)(8,0){3}{\line(1,0){3}}
\end{picture} }
\def\olk{ \begin{picture}(77,22)(0,+8.5)
     \thicklines
  \multiput(4,11)(8,0){3}{\line(1,0){3}}
\put(36,11){\circle{25}}
  \multiput(44,20)(6,3){4}{\line(1,0){3}}
  \multiput(44,2)(6,-3){4}{\line(1,0){3}}
\end{picture} }

\begin{equation}
{\rm Tr} \; \ln \left \{ 1 + G_{\beta}(\varphi_{f})
\left[ V''(\varphi_{b}) - V''(\varphi_{f}) \right] \right \} =
 \ols \! \! \! \! \! \! \! \! \! \! \! \! \! + \olr +
\olk  + \: \: \ldots \: ,
\label{graphic}
\end{equation}

\vspace{0.7 cm}




\noindent
where the dashed lines correspond to the background ``field''
$\left [ V''(\varphi_{b}) - V''(\varphi_{f}) \right ]$
and the internal lines denote
the propagator $G_{\beta}(\varphi_{f})$. The expression (\ref{graphic})
can be written
as

\begin{eqnarray}
{\rm Tr} \; \ln \left\{ 1 + G_{\beta}(\varphi_{f})
\left[ V''(\varphi_{b}) - V''(\varphi_{f}) \right] \right\} &=&
\sum_{m=1}^{+ \infty}
\frac{ (-1)^{m+1} }{m} \int d^{3} x  \left[V''(\varphi_{b}) - V''(\varphi_{f})
\right]^{m}  \times \nonumber \\
&\times &  \sum_{n= -\infty}^{+ \infty} \int \frac{ d^{3} k}
{(2 \pi)^{3}} \frac{1}{ \left[ \omega_{n}^{2} + \vec{k}^{2} +
m^{2}(\varphi_{f})
\right]^{m} } \: .
\label{traceln}
\end{eqnarray}

The sum in $m$ can be performed and one obtains

\begin{equation}
{\rm Tr} \; \ln \left\{ 1 + G_{\beta}(\varphi_{f})
\left[ V''(\varphi_{b}) - V''(\varphi_{f}) \right] \right\} =
\int d^{3} x  \sum_{n= -\infty}^{+ \infty} \int \frac{ d^{3} k}
{(2 \pi)^{3}} \ln \left[ 1 + \frac{ V''(\varphi_{b}) - V''(\varphi_{f}) }
{\omega_{n}^{2} + \vec{k}^{2} + m^{2}(\varphi_{f}) } \right] \: .
\label{endtrln}
\end{equation}

\noindent
This expression can be further simplified by means of the identity,
($E_b^2 (\vec{k})\equiv\vec{k} +m^2(\vp_b)$, and $E_f^2(\vec{k})
\equiv\vec{k} +m^2(\vp_f)$)

\begin{eqnarray}
\sum_{n= -\infty}^{+ \infty} \ln \left[ \frac{ \omega_{n}^{2} +
E_{b}^{2}(\vec{k}) }{ \omega_{n}^{2} + E_{f}^{2}(\vec{k}) } \right] &=&
\beta E_{b}(\vec{k}) + 2 \ln \left( 1 - e^{- \beta E_{b}(\vec{k}) } \right)
- \nonumber \\
&-& \beta E_{f}(\vec{k}) - 2 \ln \left( 1 - e^{- \beta E_{f}(\vec{k}) } \right)
\: .
\label{sumln}
\end{eqnarray}

\noindent
The terms proportional to $\beta$ can be renormalized. We are left with the
familiar temperature corrected effective potential (we
neglect terms coming from zero temperature quantum corrections) \cite{jackiw},

\begin{equation}
V_{\rm eff}(\varphi,T) = V(\varphi) + T \int \frac{d^{3} k}{(2 \pi)^{3}}
\ln \left( 1 - e^{- \beta \sqrt{\vec{k}^{2} + m^{2}(\varphi)}} \right)\:.
\label{Veffren}
\end{equation}

Within the thin-wall limit, using (\ref{sumln}) into (\ref{endtrln}), and
substituting into
Eq. (\ref{expratio}), we obtain, from Eq. (\ref{gamma2})
the temperature corrected barrier,

\begin{equation}
\Delta F(T) = - \frac{ 4 \pi R^{3}}{3} \Delta V_{\rm eff}(T) + 4 \pi R^{2}
\sigma (T) \: ,
\label{fielddeltaF}
\end{equation}

\noindent
where

\begin{equation}
\Delta V_{\rm eff} (T) = V(\varphi_{f}) - V(\varphi_{t}) - \frac{1}{\beta}
\int \frac{ d^{3} k}{(2 \pi)^{3}} \ln \left[ {{1-e^{-\beta E_t(\vec{k})}}\over
{1-e^{-\beta E_f(\vec{k})}}}\right ]
\label{deltaVeff}
\end{equation}

\noindent
and

\begin{equation}
\sigma (T) = \sigma_{0} + \frac{1}{4 \pi R^{2}} \int d^{3} x \frac{1}{2 \beta}
\sum_{n= - \infty}^{+ \infty} \int \frac{ d^{3} k}{(2 \pi)^{3}} \ln \left[ 1 +
\frac{V''(\varphi_{wall}) - V''(\varphi_{f}) }
{\omega_{n}^{2} + \vec{k}^{2} + m^{2}(\varphi_{f}) } \right] \: .
\label{fieldsigmaT}
\end{equation}

Expression (\ref{deltaVeff}) for the temperature corrected
false vacuum energy density
can be easily identified
with $\Delta V_{\rm eff}(T)$, as defined in Eq.
(\ref{deltaU}).
Also, expression (\ref{fieldsigmaT}) for $\sigma (T)$
can be identified with the temperature corrected
surface tension of the bubble wall, by writing it
as

\begin{equation}
\sigma (T) = \frac{1}{4 \pi R^{2} } \left[
F (\varphi_{wall},T) -
F (\varphi_{f},T) \right] \: ,
\label{defsigmaT}
\end{equation}

\noindent
with

\begin{equation}
F(\varphi,T) = - \frac{1}{\beta} \ln Z(\varphi)  \:,
\label{FphiT}
\end{equation}


\noindent
where $F(\varphi,T)$
is the free energy for the field configuration $\varphi$.
$\sigma (T)$, as given by (\ref{defsigmaT}), is then the free energy
difference (per unit area of the bubble) between the
two configurations ($\varphi_{wall}$ and $\varphi_{f}$),
which defines the surface tension. As shown in Ref. \cite{gil}, the
expressions for the free energies in
(\ref{defsigmaT}), at 1-loop order give (\ref{fieldsigmaT}).

\section{\bf EFFECTS OF INTERACTIONS ON THE DECAY RATE}

In the previous two Sections we have obtained the temperature effects on
the nucleation barrier by {\it assuming} that the system becomes trapped in a
metastable phase as the temperature drops below the critical temperature
$T_C$. This assumption is not very realistic, and was addopted so that we
could stress the main points of the calculation without having to worry
about the effects of interactions of the order parameter with other fields.
However, in models of interest within a cosmological context, such as
the standard
electroweak model and some of its extensions, the Universe becomes trapped in
a metastable phase due to the interactions of the Higgs field (or effective
scalar order parameter) with other massless fields, such as gauge bosons or
fermions. In fact, it is possible to formulate this statement in terms of
conditions that a general effective potential must satisfy, if it is to
develop a metastable phase below $T_C$
\cite{gleiser}. Basically, the condition states that the mass gap between
the symmetric and broken-symmetric phases must be large enough so that
massive fluctuations away from the symmetric minimum are suppressed. Since
the mass gap is given in terms of the (temperature dependent)
vacuum expectation value of the Higgs
field and of its coupling to other fields, the condition states that
the transition cannot be too weakly first order.

The question then is what
effective potential should be considered when calculating the decay rate. As
recently shown by E. Weinberg for the zero-temperature Coleman-Weinberg models
(which exhibit symmetry breaking only due to radiative corrections),
the effective potential relevant in the calculation of the bounce solution is
obtained by integrating over all fields {\it but} the scalar field;
the radiative
corrections coming form these fields modify the vacuum structure of the model,
making metastability possible.  In this Section, we argue that the same
procedure must be followed when calculating the nucleation rate at finite
temperature. We first review, in general terms,
how to take into account the interactions of the scalar field with other
fields in the calculation of the decay rate. (For details see, Ref.
\cite{weinb}.) Then we apply our method to
an example involving fermions, explicitly comparing our results to those
obtained from Eq. (\ref{badrate}).

\subsection{\bf General Formalism}

Let us consider a system described by a scalar field $\phi$ and a set of
fields $\xi$ (bosonic or fermionic fields) which are coupled to
$\phi$. The partition function of the system is given by

\begin{equation}
Z = \int D \phi D \xi e^{- S_{E} (\phi, \xi)} \: ,
\label{Zphixi}
\end{equation}

\noindent
where $S_{E} (\phi, \xi)$ is the euclidean action of the system and
the functional integration is carried over field configurations subject
to periodic boundary conditions, $\phi(\vec{x}, 0) = \phi(\vec{x},\beta)$,
for bosons or antiperiodic, $\psi(\vec{x},0) = - \psi(\vec{x},\beta)$,
for fermion fields.

If one integrates out the $\xi$ fields in (\ref{Zphixi}), the partition
function can be written as

\begin{equation}
Z = \int D \phi e^{- W^{\beta}(\phi)}  \: ,
\label{ZWphi}
\end{equation}

\noindent
where

\begin{equation}
W^{\beta}(\phi) = - \ln \int D \xi e^{- S_{E}(\phi, \xi)} \: .
\label{Wfunc}
\end{equation}

\noindent
$W^{\beta}(\phi)$ can be viewed as an effective action for the scalar
field $\phi$, where only $\xi$-loop terms are included. Note that these
$\xi$-loop terms introduce finite temperature corrections in (\ref{Wfunc}).
$S_{E}(\phi, \xi)$ in (\ref{Zphixi}) and (\ref{Wfunc})
includes renormalization counter-terms
and, if one or more of the $\xi$-fields
is a gauge field, $S_E(\phi, \xi)$
also includes the gauge fixings and the corresponding
ghost terms.

The procedure is now, in principle, straightforward. We evaluate
the partition function in Eq. (\ref{ZWphi}) semiclassically by expanding the
effective action $W^{\beta}(\phi)$ around its extremum configuration, which
will be the bounce. Note that the bounce will include the radiative
corrections coming from the fields that couple to $\phi$, {\it
but not from $\phi$
itself.}
The determinantal prefactor  can be evaluated as before, by considering the
negative and zero eigenvalues separately from the positive eigenvalues.
However, as pointed out by Weinberg \cite{weinb}, $W^{\beta}(\phi)$ cannot
always be obtained in closed form,
being in general a nonlocal functional. He proposes to resolve this difficulty
by considering
a local action $W_{0}(\phi)$ which is
close enough to the original one. From
a derivative expansion of $W^{\beta}(\phi)$, $W_0(\phi)$
is found to be

\begin{equation}
W_{0}(\phi) = \int d^{4} x_E~ \left[ \frac{1}{2} (\partial_{\mu}\phi)^{2} +
\hat{V}_{1-loop}(\phi) \right] \: ,
\label{Wphi0}
\end{equation}

\noindent
where $ \hat{V}_{1-loop}(\phi)$ includes the tree level potential
$V(\phi)$ and the 1-loop contributions coming {\it only} from the $\xi$-field
integration. The bounce solution can then be obtained from
(\ref{Wphi0}). $W_0(\phi)$ should be a good approximation to $W^{\beta}(\phi)$
as long
as the typical interaction length scale set by the field(s) $\xi$ is shorter
than the scale for the nonlocality of $W^{\beta}(\phi)$.

In what follows we will apply the above formalism to a specific example of
a scalar field coupled to fermions. For large enough Yukawa couplings, this
model has been shown to satisfy the conditions for metastability specified
in Ref.
\ref{gleiser}.

\subsection{\bf Application: Real scalar
field coupled
to fermion fields}

Consider a model of a real scalar field $\phi$
coupled to fermion fields with a lagrangian density

\begin{equation}
{\cal L} = \frac{1}{2} (\partial_{\mu} \phi)^{2} - V(\phi) -
g \phi \bar{\psi} \psi + i \bar{\psi} \not{\! \partial} \psi \: ,
\label{Lphipsi}
\end{equation}

\noindent
with $V(\phi)$ given by

\begin{equation}
V(\phi) = \frac{\lambda}{4 !} \phi^{4} - \frac{\alpha\mu}{3 !} \phi^{3} +
\frac{\mu^{2}}{2} \phi^{2} \: ,
\label{Vphi}
\end{equation}

\noindent
where $\lambda$ and $\alpha$ are positive, dimensionless constants
and $\mu^{2} > 0$ is a (mass)$^{2}$ parameter. A large enough
coupling to fermions
guarantees that a metastable phase is possible as the system is cooled below
$T_C$; the high temperature minimum of
the 1-loop effective potential,
$\langle \varphi \rangle_{T}\simeq \frac{\alpha}{\lambda + 4 g^{2}} \mu$ ,
lies to the left of the maximum of the potential at $T=T_C$, and
thermal fluctuations away from the symmetric minimum are suppressed. Numerical
values for the couplings satisfying these conditions can be found in Ref.
\ref{gleiser}.

The partition function for this model is given by

\begin{equation}
Z = \int D \phi D \psi D \bar{\psi} e^{ - \int_{0}^{\beta} d \tau \int
d^{3} x {\cal L} ( \phi, \psi, \bar{\psi})} \: .
\label{Zphipsi}
\end{equation}

\noindent
As the fermion fields appear quadratically in (\ref{Zphipsi}), they can be
integrated out, giving

\begin{equation}
Z= \int D \phi e^{- W_0(\phi)} \: ,
\label{ZWscalar}
\end{equation}

\noindent
where

\begin{equation}
W_0(\phi) = \int_{0}^{\beta} d \tau \int d^{3} x  \left[ \frac{1}{2}
(\partial_{\tau} \phi)^{2} + \frac{1}{2} ( \vec{\nabla} \phi)^{2} +
V(\phi) \right] - {\rm Tr} \ln ( - \not{\! \partial}  - i g \phi)_{\beta} \: ,
\label{Wbetaphi}
\end{equation}

\noindent
and

\begin{equation}
{\rm Tr} \ln  ( - \not{\! \partial}  - i g \phi)_{\beta} = \ln \det
( - \not{\! \partial}  - i g \phi)_{\beta} \: .
\label{trln}
\end{equation}

Thus, $W_0(\phi)$ can be written as

\begin{equation}
W_0(\phi) = \int_{0}^{\beta} d \tau \int d^{3} x  \left[ \frac{1}{2}
(\partial_{\tau} \phi)^{2} + \frac{1}{2} ( \vec{\nabla} \phi)^{2} +
{\hat V}_{\psi}(\phi,T) \right] \: ,
\label{Wbeta}
\end{equation}
where, for sufficiently smooth fields (see Appendix B),
the effective potential obtained after integrating
over the fermions and renormalizing is (we will drop all zero temperature
quantum corrections)

\begin{equation}
\hat{V}_{\psi}(\p,T) = V(\p)
-
4 T \int \frac{d^{3} k}{(2 \pi)^{3}} \ln \left( 1 + e^{- \beta
\sqrt{ \vec{k}^{2} + g^{2} \p^{2}} } \right) \: ,
\label{Vbetaphi}
\end{equation}

\noindent
The temperature dependent term accounts for finite temperature corrections
coming from fermion loops. $\hat{V}_{\psi}(\p,T)$ is the potential we should
use to compute the bounce. Note that, neglecting the $T=0$ quantum corrections,
the high-temperature limit of $\hat{V}_{\psi}(\p,T)$ is approximately

\begin{equation}
\hat{V}_{\psi}(\p,T) \simeq V(\p) + \frac{T^{2}}{12} g^{2}
\p^{2} \:,
\label{VphiT}
\end{equation}

\noindent
so that for $(T/\mu)^2> (9\alpha^2/\lambda-24)/4g^2$, a condition which
is easily satisfied for reasonable values of the couplings,
the high-temperature minimum is
$\langle \varphi \rangle_{T} \simeq 0$.

Once we have the action $W_0(\phi)$, the bounce is obtained as a
solution of

\begin{equation}
\frac{\delta W_0(\phi)}{\delta \phi} |_{\phi = \varphi_{b}} = 0 \: .
\label{Wvariat}
\end{equation}

\noindent
Thus, for a static, spherically symmetric configuration, the bounce
configuration $\vp_b(r)$ will be a solution of

\begin{equation}
\frac{d^{2} \phi}{d r^{2}} + \frac{2}{r} \frac{d \phi}{d r} =
\hat{V}_{\psi}' (\phi,T)  \: ,
\label{bubbsphe}
\end{equation}

\noindent
with boundary conditions, $\lim_{r \rightarrow \infty} \varphi_{b}(r)
= \varphi_{f}\simeq 0$ and $\frac{d \varphi_{b}}{d r} |_{r =0} = 0$. (From
now on $\varphi_{f}$ and $\varphi_{t}$
should be understood as being the minima of (\ref{Vbetaphi})
and not the minima
of the tree-level potential $V(\p)$.)

The procedure is now identical to that of Sections 2 and 3. Having a bounce
solution we can evaluate the partition function
written in Eq. (\ref{ZWscalar}) semiclassically,
exactly as was done in
Eqs.
(\ref{Zphib}) and (\ref{Zphif}), by expanding around $\varphi_f$ and
$\varphi_b$. We then obtain, from Eqs.
(\ref{energy}) and (\ref{gamma}),
the nucleation rate,

\begin{equation}
{\cal R}= - \frac{|E_- |}{\pi} {\rm Im} \left[ \frac{ \det ( -\Box_{E} +
m_{\beta}^{2} (\varphi_{b}))_{\beta}}{ \det ( -\Box_{E} + m_{\beta}^{2}
(\varphi_{f}))_{\beta}}
\right]^{-\frac{1}{2}}
e^{- \Delta W_0} \: ,
\label{gammaW}
\end{equation}

\noindent
where $m_{\beta}^{2}(\varphi) = \frac{d^{2} \hat{V}_{\psi}(\phi)}{
d \phi^{2}}|_{\phi = \varphi}$, with $\hat{V}_{\psi}(\phi)$ obtained
above.
$ \Delta W_0$ is given by

\begin{equation}
\Delta W_0 = W_0(\varphi_{b}) - W_0(\varphi_{f}) \: ,
\label{deltaW}
\end{equation}

\noindent
where $ W_0(\p)$ was defined in Eq. (\ref{Wbeta}).

In order to proceed, we must rewrite the determinantal prefactor
explicitly isolating the negative and zero modes from the positive modes. This
is done following the same steps of Section 3, although now we must handle
the fermionic contribution to the determinants. The details of the
perturbative expansion for the
fermionic determinantal prefactor are given in Appendix B. We can then
write the nucleation rate per unit volume as

\begin{equation}
\Gamma =
\frac{T^4}{\pi} \frac{ \frac{|E_{-}|}{2 T}}{ \sin \left( \frac{
|E_{-}|}{2 T} \right)} \left[ \frac{\Delta W_0}{2 \pi}
\right]^{\frac{3}{2}}
\exp \left[ - \frac{ \Delta F(T) }{T} \right] \: ,
\label{endgammaW}
\end{equation}

\noindent
where $\Delta F(T)$, the bubble activation free energy in the 1-loop
approximation, is given by

\begin{equation}
\Delta F(T) = \int d^{3} x \left\{{1\over 2}\left (\nabla\vp_b\right )^2+
\hat{V}_{\psi}(\varphi_{b}) -
\hat{V}_{\psi}(\varphi_{f}) + \frac{1}{2 \beta} \sum_{n= - \infty}^{
+ \infty} \int \frac{ d^{3}k}{(2 \pi)^{3}} \ln \left[
1 + \frac{ m_{\beta}^{2}(\varphi_{b}) -   m_{\beta}^{2}(\varphi_{f})}{
\omega_{n}^{2} + \vec{k}^{2} +  m_{\beta}^{2}(\varphi_{f})}
\right] \right\} \: .
\label{DFT}
\end{equation}

\noindent
As usual, the sum over $n$ can be performed and we get,

\begin{equation}
\Delta F(T)=\int d^3x \left [{1\over 2}\left (\nabla\vp_b\right )^2+
\hat{V}_{{\rm eff}}(\vp_b,T) -\hat{V}_{{\rm eff}}(
\vp_f,T)\right ]  \:,
\label{Veff}
\end{equation}

\noindent
where the effective potential $\hat{V}_{{\rm eff}}(\p,T)$, is given by
(neglecting zero temperature quantum corrections)

\begin{eqnarray}
\hat{V}_{{\rm eff}}(\phi,T) &=& V(\p) -
4 T \int \frac{d^{3}k}{(2 \pi)^{3}}
\ln \left( 1 + e^{- \beta \sqrt{ \vec{k}^{2} + g^{2} \p^{2}}}
\right) + \nonumber \\
&+& T \int \frac{d^{3}k}{(2 \pi)^{3}}
\ln \left( 1 - e^{- \beta \sqrt{ \vec{k}^{2} + m_{\beta}^{2}(\p)}}
\right) \:,
\label{hatVeff}
\end{eqnarray}
\noindent
where $V(\p)$ is the tree level potential (\ref{Vphi}) and the mass term
appearing in the scalar loop contribution is
$m_{\beta}^{2} (\p) \simeq V''(\p) +
\frac{T^{2}}{6} g^{2}$ in leading order in the fermion loops. It is
instructive to contrast this result with that obtained for self-coupled
scalars, Eq. (\ref{Veffren}).  The coupling to fermions modifies the scalar
mass propagating in the loops. This effect naturally improves the infrared
behavior of the theory, and can be of importance in weak first-order
transitions. We will say more about this later.
Note also a
crucial difference between this expression for the free energy barrier
and the expression for the free energy barrier in Eq. (\ref{badrate}): Here,
the bounce is obtained with the effective
potential that does {\rm not} include the
corrections coming from scalar loops. The corrections from scalar loops
which appear in the last term of Eq. (\ref{hatVeff}) are thermally
induced fluctuations about
the bounce solution computed with $\hat V_{\psi}(\p,T)$. In the usual
expression
for the nucleation barrier the bounce is obtained from the full effective
potential including the scalar loops. The two expressions are definetely
not equivalent, even though for small scalar self-couplings the differences are
negligible. In order to illustrate the differences let us look at a
specific example.

In Fig. 2 we contrast the two approaches by comparing the nucleation
barriers as a function of the temperature for a fixed set of coupling
constants. The barriers in the figure were obtained by a numerical
integration of the bounce equation including the relevant loops according
to each approach.
For clarity let us call the nucleation barrier obtained in the usual
approach, {\it i.e.}, by including the scalar loops in the bounce calculation,
the scalar barrier. The nucleation barrier obtained without including the
scalar loops we call the fermionic barrier.
In Fig. 2 we take $\lambda =1.0,~
\alpha=2.0$, and $g^2=0.5$. Since $\lambda$ controls the strength of the
scalar corrections, we expect the differences between the two barriers to
be noticeable. We find that this is indeed the case, noting that as
we approach the critical temperature (that is, as we move closer
to the thin-wall limit) the differences between the two barriers increase,
with the scalar barrier {\it always larger} than the fermionic barrier. This is
precisely what one expects if the scalar corrections are entropic corrections
to the nucleation barrier. Thus, the nucleation barrier used in expression Eq.
(\ref{badrate}) is overestimated for large enough scalar corrections.

Finally, we point out two additional differences between the results.
First note that the contribution from the zero modes
to the prefactor depends on $\Delta W_0(\phi)$, as opposed
to $\Delta F(T)$. This could be important for weak transitions in which the
prefactor may play a relevant r\^ole. Most importantly,
the expression for $\hat{V}_{\rm eff}(\p,T)$, Eq.~(\ref{hatVeff}),
differs from the
usual 1-loop finite temperature effective potential
by the mass term for the scalar field loops,
$m_{\beta}^{2}(\p) = \hat{V}_{\psi}''(\p)$. Since we have
used the stationary points of $W_0(\phi)$, Eq. (\ref{Wbeta}),
as opposed to the stationary points of $S_E(\p,\psi)$, as
the effective ``background'' fields
in the saddle-point  evaluation of the partition function,
the scalar field propagator carries the finite
temperature mass $m_{\beta}^{2}(\p)$. The propagator is dressed
by the quantum corrections due to fermion loops. In the usual 1-loop finite
temperature effective potential, the stationary points are obtained from
the tree level action, with mass term for scalar loops,
$m_{0}^{2}(\p) = V''(\p)$. This results in the usual negative mass
terms related to the change in convexity of the effective potential between
the inflection points, and,
in the case of very shallow potentials, in bad infrared behavior near $\vp_f$.
The incorporation of the fermionic corrections to the scalar propagator,
which is demanded by our method of calculation atenuates these problems. In the
example above, the scalar mass gets dressed by fermionic loops, being given by
$m_{\beta}^{2}(\p) \simeq V''(\p) + \frac{1}{6} g^{2} T^{2}$,
where $V(\p)$ is the tree-level potential (\ref{Vphi}). The
temperature term in $m_{\beta}$ works as the infrared regulator for
small values of $m_{0}^{2}(\p) = V''(\p)$. This result is
independent of the particular model studied. Similar conclusions have been
obtained in Ref. \cite{buch} for scalar electrodynamics.

\section{\bf Conclusions}

In this paper we examined in some detail the computation of false vacuum
decay rates at finite temperatures in the regime in which quantum fluctuations
are negligibly small compared to thermal fluctuations.
We have shown that temperature
corrections to the nucleation barrier can be obtained from a saddle-point
evaluation of the partition function in a dilute gas approximation. In fact,
the temperature corrections are simply due to the positive eigenvalues from
stable fluctuations around the critical bubble. That is, they are the
entropic contributions due to thermally induced deformations on the bubble.

Even though this result has been known in classical
statistical mechanics for more than two decades \cite{langer67},
we believe that a
consistent treatment within field theory is still lacking. Although we left
many questions unanswered, we hope to have clarified some of the issues
involved in the calculation of finite-temperature decay rates. Of particular
importance is the fact that the bounce is {\it not}
obtained from the full 1-loop
corrected effective potential, but from the potential excluding the scalar
loops. Thus, for a self-interacting scalar, the bounce is obtained from
the tree-level potential.
The full finite temperature potential appears in the exponent
only after properly accounting for the positive eigenvalues of the
determinantal prefactor. That is, the scalar contributions account for
entropic corrections to the nucleation barrier.
We obtained a temperature corrected nucleation
barrier which can differ from the usual result. We showed
this to be particularly true for sufficiently large scalar self-couplings
in the vicinity of the critical temperature for the transition.

Also, we found that the interaction with other
fields gives rise to a potential which is better behaved in the infrared.
(See also Ref. \cite{buch}.)
This result is the finite-temperature equivalent to what E. Weinberg found
for the zero-temperature case, once the integration over the other fields
is performed \cite{weinb}.

The reader may be wondering if our results
will have any consequences to current work on the
electroweak phase transition. The answer depends on the Higgs mass. For a
sufficiently light
Higgs it is consistent to neglect the contribution from scalar loops to
the effective potential. In this case, the usual estimate for
the nucleation barrier is a valid approximation. However, the
situation may change
for a heavier Higgs. Given that the experimental lower bound on the Higgs
mass is now above 60 GeV, we believe it worthwhile
to study this question in more detail, keeping in mind that
the transition becomes weaker as the Higgs mass increases.

\acknowledgments

We would like to thank A. Linde for many important discussions on these and
related issues.
We would like to thank the Institute for Theoretical Physics in Santa Barbara
where, during the program on Cosmological Phase Transitions, this work begun.
At ITP this work was supported in part by a National Science
Foundation grant No.
PHY89-04035 at ITP. (MG) was supported in part by a National Science
Foundation grant No. PHYS-9204726.
(GCM) aknowledges
financial support from FAPESP (S\~ao
Paulo, Brazil) and (ROR) from
Conselho Nacional
de Desenvolvimento Cient\'{\i}fico e Tecnol\'ogico - CNPq (Brazil).

\vspace{0.5in}
\centerline{\bf Figure Captions}

\vspace{0.3in}

\noindent
{\bf Figure 1:} A typical asymmetric double-well potential.\\

\noindent
{\bf Figure 2:} A comparison of the nucleation barrier as a
function of temperature,
in units of mass parameter
$\mu$, obtained by including
(stars) and excluding (dots) scalar loops in the computation of the
bounce. The parameters in the tree-level potential
are $\lambda=1.0,~\alpha=2.0,~g^2=0.5$.\\

\noindent
{\bf Figure 3:} A comparison of the terms $g^2\p^2(r)$
and $g~d\p(r)/dr$ appearing in
Appendix B. The parameters in the tree-level potential are $\lambda=1,~\alpha=
0.56,~g^2=0.5$.\\


\appendix
\section{}

Let us show that the determinant ratio of Eq. (\ref{ratio}),
denoted here as $R$,
gives (\ref{eigenratio}).
Separating the negative and zero eingevalues in the denominator of Eq.
(\ref{ratio}), one can write

\begin{equation}
R = \exp \left\{ \frac{1}{2} {\rm ln} \left[ \frac{
\prod_{n= -\infty}^{+ \infty} \prod_{i} \left( \omega_{n}^{2} + E_{f}^{2}(i)
\right)}{\prod_{n= -\infty}^{+ \infty} (\omega_{n}^{2} + E_{-}^{2})
(\omega_{n}^{2} + E_{0}^{2})^{3} \prod_{j} ' \left( \omega_{n}^{2} +
E_{b}^{2}(j) \right)} \right] \right\} \: ,
\label{Aratio}
\end{equation}

\noindent
where the prime in $\prod_{j}$ means that the negative eigenvalue, $E_{-}^{2}$,
and the three zero eigenvalues, $E_{0}^{2}$, are now excluded from the product.
The term for $n=0$ in $(\omega_{n}^{2} + E_{0}^{2})$, can be handled as in
Ref. \cite{coleman}, resulting in the factor ${\cal V} \left[\frac{
\Delta E}{2 \pi T} \right]^{\frac{3}{2}}$ in Eq. (\ref{eigenratio}).
Separating the $n=0$
modes both in the numerator and the denominator of (\ref{Aratio}), and using
the identity (\ref{idenpi}),
we get,

\begin{eqnarray}
R &=& {\cal V} \left[ \frac{ \Delta E}{2 \pi T} \right]^{\frac{3}{2}}
\exp\left\{\left(-4 + \sum_{i} - \sum_{j}\; ' \right) {\rm ln}
\prod_{n=1}^{+\infty}
\omega_{n}^{2} - {\rm ln} \left( E_{-}^{2} \right)^{1/2} - {\rm ln} \left[
\frac{\sin(\frac{\beta}{2} |E_{-}|)}{\frac{\beta}{2} |E_{-}|} \right]
+ \right. \nonumber \\
&+& \left. \left(\sum_{j}\; '- \sum_{i}\right) {\rm ln} \beta +
\sum_{i} \left[ \frac{\beta}{2} E_{f}(i) + {\rm ln} \left( 1 - e^{- \beta
E_{f}(i)} \right) \right] + \right. \nonumber \\
&-& \left. \sum_{j} \; ' \left[ \frac{\beta}{2} E_{b}(j) + {\rm ln}
\left( 1 - e^{- \beta E_{b}(j)} \right) \right] \right\} \: .
\label{Aexplicity}
\end{eqnarray}

\noindent
In the above expression we used that the negative eigenvalue can be written as
$(E_{-}^{2})^{\frac{1}{2}} = i |E_{-}|$. Remembering that $\sum_{j} '$
has four eigenvalues less than $\sum_{i}$, we can write

\begin{eqnarray}
R &=& {\cal V} \left[ \frac{ \Delta E}{2 \pi T} \right]^{\frac{3}{2}}
\exp \left\{ -4 {\rm ln} \beta - {\rm ln} \left( E_{-}^{2} \right)^{1/2} -
{\rm ln} \left[\frac{\sin(\frac{\beta}{2} |E_{-}|)}{\frac{\beta}{2}
|E_{-}|} \right] + \right. \nonumber \\
&+& \left.
\sum_{i} \left[ \frac{\beta}{2} E_{f}(i) + {\rm ln} \left( 1 - e^{- \beta
E_{f}(i)} \right) \right]
-  \sum_{j} \; ' \left[ \frac{\beta}{2} E_{b}(j) + {\rm ln}
\left( 1 - e^{- \beta E_{b}(j)} \right) \right] \right\}
\label{Aendexplicitly}
\end{eqnarray}

\noindent
which reduces to Eq. (\ref{eigenratio}).


\section{}

In (\ref{gammaW}), the exponential term
$\Delta W_0$ can be written as (from (\ref{deltaW}) and (\ref{Wbetaphi}))

\vspace{0.5cm}

\begin{equation}
\Delta W_0 = \left[ S_{E}(\varphi_{b}) - {\rm Tr} \ln
(- \not{\! \partial} - i g\varphi_{b})_{\beta} \right] -
\left[ S_{E}(\varphi_{f}) - {\rm Tr} \ln
(- \not{\! \partial} - i g\varphi_{f})_{\beta} \right] \: ,
\label{BdeltaW}
\end{equation}

\noindent
where $S_{E}(\p) = \int_{0}^{\beta} d \tau \int d^{3} x \left[
\frac{1}{2} (\partial_{\mu} \p)^{2} + V(\p) \right]$, is
the classical action for the scalar field as in (\ref{Lphipsi}). The
exponential
in (\ref{gammaW}) can be written then as

\begin{equation}
e^{- \Delta W_0} = \left[ \frac{ \det ( - \not{\! \partial} - i g
\varphi_{b})_{\beta} }{ \det ( - \not{\! \partial} - i g \varphi_{f}
)_{\beta} }
\right] e^{- \Delta S} \: ,
\label{BexpW}
\end{equation}

\noindent
with $\Delta S = S_{E}(\varphi_{b}) - S_{E}(\varphi_{f})$.

If one uses the identity (which follows from charge-conjugation invariance):

\begin{eqnarray}
\left[ \det ( - \not{\! \partial} - i g \varphi) \right]^{2} &=&
\det ( - \not{\! \partial} -i g \varphi).\det( - \not{\! \partial} +
i g \varphi)
= \nonumber \\
&=& \det \left[ ( -\Box_{E} + g^{2} \varphi^{2} ) 1_{4 \times 4} -
i g \gamma_{E}^{\mu} \partial_{\mu} \varphi \right] \: ,
\label{Bident}
\end{eqnarray}

\noindent
where $1_{4 \times 4}$ is the $4 \times 4$ unit matrix, then, for $\varphi=
\varphi_{f}$, one gets

\begin{equation}
\det ( - \not{ \! \partial} - i g \varphi_{f})_{\beta} =
\left[ \det ( - \Box_{E} + g^{2} \varphi_{f}^{2} )_{\beta} \right]^{
\frac{1}{2}} \: .
\label{Bdetf}
\end{equation}

\noindent
For the determinant involving the spherically symmetric bounce $\varphi_{b}$,
the Dirac matrix
$\gamma_{E}^{\mu}$ in (\ref{Bident}) is radial and one can write

\begin{equation}
\gamma_{r} = i \left(
\begin{array}{cc}
1_{2 \times 2} & 0 \\
0 & - 1_{2 \times 2}
\end{array}
\right) \: ,
\label{Bmatrix}
\end{equation}

\noindent
where $1_{2 \times 2}$ denotes a $ 2 \times 2$ unit matrix. Then, for
$\varphi_{b} = \varphi_{b}(r)$,
$\det ( - \not{\! \partial} - ig \varphi_{b}(r) )_{\beta}$ can be written as

\begin{equation}
\det ( - \not{\! \partial} - ig \varphi_{b}(r) )_{\beta} =
\det \hat{ \Omega}^{(+)}(\varphi_{b}) . \det \hat{\Omega}^{(-)}(\varphi_{b})
\: ,
\label{Bdetb}
\end{equation}

\noindent
where

\begin{equation}
\hat{\Omega}^{(\pm)}(\varphi_{b}) = - \Box_{E} + g^{2} \varphi_{b}^{2} \pm
g \frac{\partial \varphi_{b}}{\partial r} \: .
\label{Bomega}
\end{equation}

\noindent
Therefore, the determinants in (\ref{BexpW}) can be written as
(using that $\ln \det \hat{M} = \break {\rm Tr} \ln \hat{M}$ )

\begin{eqnarray}
\frac{ \det ( - \not{\! \partial} - i g \varphi_{b} )_{\beta} }{ \det (
- \not{\! \partial} - i g \varphi_{f} )_{\beta} } &=&
\exp \left\{ {\rm Tr} \ln ( - \not{\! \partial} - i g \varphi_{b} )_{\beta} -
{\rm Tr} \ln ( - \not{\! \partial} - i g \varphi_{f} )_{\beta} \right\}
= \nonumber \\
&=& \exp \left\{ {\rm Tr} \ln \left[ -\Box_{E} + g^{2} \varphi_{b}^{2} +
g \frac{ \partial \varphi_{b}}{\partial r} \right]_{\beta} +
{\rm Tr} \ln \left[ - \Box_{E} + g^{2} \varphi_{b}^{2} - g \frac{ \partial
\varphi_{b}}{\partial r} \right]_{\beta} - \right. \nonumber \\
&-& \left. 2 {\rm Tr} \ln \left[ - \Box_{E} + g^{2}
\varphi_{f}^{2} \right]_{\beta}
\right\} \: .
\label{Bdetbf}
\end{eqnarray}

\noindent
As in (\ref{expratio}), one can write (\ref{Bdetbf}) as

\begin{eqnarray}
\frac{ \det ( - \not{\! \partial} - i g \varphi_{b} )_{\beta} }{ \det (
- \not{\! \partial} - i g \varphi_{f} )_{\beta} } &=&
\exp \left\{ {\rm Tr} \ln \left[ 1 + S_{\beta}(\varphi_{f}) \left[ g^{2}
( \varphi_{b}^{2}
- \varphi_{f}^{2} ) + g \frac{ \partial \varphi_{b}}{\partial r} \right]
\right]
+ \right. \nonumber \\
&+& \left. {\rm Tr} \ln \left[ 1 + S_{\beta}(\varphi_{f}) \left[ g^{2} (
\varphi_{b}^{2}
- \varphi_{f}^{2} ) - g \frac{ \partial \varphi_{b}}{\partial r} \right]
\right]
\right\} \: ,
\label{Bexpratio}
\end{eqnarray}

\noindent
where, in analogy to Eq. (\ref{expratio}), we introduce the propagator

\begin{equation}
S_{\beta}(\phi_{f}) = \frac{1}{ -\Box_{E} + g^{2} \varphi_{f}^{2} }
\label{BSphif}
\end{equation}

\noindent
The argument of the exponent
in the rhs of (\ref{Bexpratio}) can be written as a series,
analogously to
Eq. (\ref{traceln}):

\begin{eqnarray}
{\rm Tr} \ln \left[ 1 + S_{\beta}(\varphi_{f}) \left[ g^{2} ( \varphi_{b}^{2}
- \varphi_{f}^{2} ) \pm g \frac{ \partial \varphi_{b}}{\partial r} \right]
\right]
&=& \sum_{m= 1}^{+ \infty} \frac{(-1)^{m+1}}{m} \int d^{3}x \left[ g^{2}
( \varphi_{b}^{2}
- \varphi_{f}^{2} ) \pm g \frac{ \partial \varphi_{b}}{\partial r} \right]^{m}
\times \nonumber \\
&\times & \sum_{n= -\infty}^{+ \infty} \int
\frac{d^{3} k}{ (2 \pi)^{3}} \frac{1}{ \left[ \bar{\omega}_{n}^{2} +
\vec{k}^{2}
+ g^{2} \varphi_{f}^{2} \right]^{m} } \: ,
\label{Btraceln}
\end{eqnarray}

\noindent
where $\bar{\omega}_{n} = \frac{(2 n + 1) \pi}{\beta}$.
As before, (\ref{Btraceln}) can be expressed as a graphic expansion
similar to (\ref{graphic}),
with the
propagators $G_{\beta}(\varphi_{f})$ replaced now by $S_{\beta}(\varphi_{f})$
and the external lines given by $g^{2} ( \varphi_{b}^{2}
- \varphi_{f}^{2} ) + g \frac{ \partial \varphi_{b}}{\partial r}$ or
$ g^{2} ( \varphi_{b}^{2}
- \varphi_{f}^{2} ) - g \frac{ \partial \varphi_{b}}{\partial r}$ .

The determinant factor in (\ref{gammaW}), coming from the functional
integration of the scalar field, can be evaluated by the same methods
of Sec. 3. In (\ref{gammaW}), the determinant term $\det [ -\Box_{E} +
m_{\beta}^{2} (\varphi_{b}) ]_{\beta}$, with $m_{\beta}^{2} (\varphi_{b})
= \hat{V}_{\psi}''(\varphi_{b})$, has a negative eigenvalue, $E_{-}^{2}$,
associated with the instability of the critical bubble, and the three
zero eigenvalues, associated with the translational invariance of the
bubble. These eigenvalues can be handled as usual, giving the preexponential
term in (\ref{endgammaW}). The part of the determinant
involving the positive eigenvalues can be written as an expansion exactly
as in (\ref{traceln}),

\begin{equation}
\left[ \frac{ \det' ( -\Box_{E} +\hat{V}_{\psi}''(\varphi_{b}))_{\beta}}
{ \det ( -\Box_{E} + \hat{V}_{\psi}''(\varphi_{f}))_{\beta}}
\right]^{- \frac{1}{2}}
= \exp \left\{ - \frac{1}{2} {\rm Tr} \;
\ln \Bigl [ 1 + \hat{G}_{\beta}(\varphi_{f})
\left[ \hat{V}_{\psi}''(\varphi_{b}) - \hat{V}_{\psi}''(\varphi_{f})
\right] \Bigr ] \right\} \: ,
\label{Bdetbos}
\end{equation}

\noindent
with $\hat{G}_{\beta} (\varphi_{f}) = \frac{1}{- \Box_{E} + m_{\beta}^{2}
(\varphi_{f})}$
and

\begin{eqnarray}
{\rm Tr} \; \ln \left\{ 1 + \hat{G}_{\beta}(\varphi_{f})
\left[ \hat{V}_{\psi}''(\varphi_{b}) - \hat{V}_{\psi}''(\varphi_{f})
\right] \right\} &=& \sum_{m=1}^{+ \infty}
\frac{ (-1)^{m+1} }{m} \int d^{3} x  \left[\hat{V}_{\psi}''(\varphi_{b})
- \hat{V}_{\psi}''(\varphi_{f})
\right]^{m}  \times \nonumber \\
&\times &  \sum_{n= -\infty}^{+ \infty} \int \frac{ d^{3} k}
{(2 \pi)^{3}} \frac{1}{ \left[ \omega_{n}^{2} + \vec{k}^{2} + m_{\beta}^{2}
(\varphi_{f})
\right]^{m} } \: .
\label{BtrlnW}
\end{eqnarray}

\noindent
The sum in $m$ in both (\ref{Btraceln}) and (\ref{BtrlnW}) can be performed
as in Eq. (\ref{traceln})).
Therefore, from Eqs. (\ref{BexpW}), (\ref{Btraceln}) and (\ref{BtrlnW}),
we can write the relevant part of
Eq. (\ref{gammaW}) as

\begin{eqnarray}
\lefteqn{ \left[ \frac{ \det ( -\Box_{E} +
\hat{V}_{\psi}''(\varphi_{b}))_{\beta}}
{ \det ( -\Box_{E} + \hat{V}_{\psi}''(\varphi_{f}))_{\beta}}
\right]^{- \frac{1}{2}} e^{- \Delta W_0} =
\left[ \frac{ \det ( -\Box_{E} +\hat{V}_{\psi}''(\varphi_{b}))_{\beta}}
{ \det ( -\Box_{E} + \hat{V}_{\psi}''(\varphi_{f}))_{\beta}}
\right]^{- \frac{1}{2}} \: \:
\frac{ \det ( - \not{\! \partial} - i g \varphi_{b} )_{\beta} }{ \det (
- \not{\! \partial} - i g \varphi_{f} )_{\beta} } \: \: e^{-\Delta S} =}
\nonumber \\
& & = {\cal V}\frac{T^{4}}{i |E_{-}| } \frac{\beta \frac{|E_{-}|}{2}}{
\sin \left( \beta \frac{|E_{-}|}{2} \right)} \left[ \frac{\Delta W_0}
{2 \pi }
\right]^{\frac{3}{2}} \exp \left\{ - \Delta S + \int d^{3} x
\sum_{n= -\infty}^{+ \infty} \int \frac{d^3 k}{(2 \pi)^3} \left[
\ln \left( 1 + \frac{g^2 (\vp_{b}^{2} - \vp_{f}^{2}) +
g \frac{\partial \vp_{b}}{\partial r}}{\bar{\omega}_{n}^{2} + \vec{k}^2 +
g^2 \vp_{f}^{2}} \right) +  \right. \right. \nonumber \\
& & + \left. \left. \ln \left( 1 + \frac{g^2 (\vp_{b}^{2} -
\vp_{f}^{2}) -
g \frac{\partial \vp_{b}}{\partial r}}{\bar{\omega}_{n}^{2} + \vec{k}^2 +
g^2 \vp_{f}^{2}} \right) -
\frac{1}{2} \ln \left( 1 + \frac{m_{\beta}^2 (\vp_{b}) -
m_{\beta}^{2} (\vp_{f}^{2})
}{\omega_{n}^{2} + \vec{k}^2 +
m_{\beta}^{2}( \vp_{f}^{2})} \right) \right] \right\} \: ,
\label{Bdeterminantal}
\end{eqnarray}

\noindent
where $\Delta W_0 $ is given by

\begin{eqnarray}
\Delta W_0 &=& \Delta S - \int d^{3} x
\sum_{n= -\infty}^{+ \infty} \int \frac{d^3 k}{(2 \pi)^3} \left[
\ln \left( 1 + \frac{g^2 (\vp_{b}^{2} - \vp_{f}^{2}) +
g \frac{\partial \vp_{b}}{\partial r}}{\bar{\omega}_{n}^{2} + \vec{k}^2 +
g^2 \vp_{f}^{2}} \right) + \right. \nonumber \\
&+& \left. \ln \left( 1 + \frac{g^2 (\vp_{b}^{2} -
\vp_{f}^{2}) -
g \frac{\partial \vp_{b}}{\partial r}}{\bar{\omega}_{n}^{2} + \vec{k}^2 +
g^2 \vp_{f}^{2}} \right)  \right] \: .
\label{finaldeltaW}
\end{eqnarray}

Apart from the derivative terms $\partial\vp_b/\partial r$, the momentum
integral reproduces the finite temperature corrections to the the tree-level
potential appearing in $\Delta S$. When we wrote  the expression for $\Delta
F(T)$ in Eq. (\ref{DFT}), these terms were not included in the effective
potential $\hat{V}_{\rm eff}(\p ,T)$. There are two reasons for negleting
this term. First, due to the graphic expansion we used for the determinants,
it is easy to see that at least at the tadpole level, their contribution
cancels. Since the tadpole gives the dominant temperature contribution to
the potential, terms that depend on  $\partial\vp_b/\partial r$ will be
sub-dominant. Second, it is possible to explicitly compare the terms
$g^2\vp_b^2$ and $g \partial\vp_b/\partial r$, by obtaining $\vp_b(r)$
numerically. We have performed this comparison for the same set of parameters
used in Figs. 2 and 3, and convinced ourselves
that the derivative term will indeed be sub-dominant.
A typical example is shown in
Fig. 4. Thus, neglecting the term  $\partial\vp_b/\partial r$,
we can use Eqs. (\ref{Bdeterminantal}) and (\ref{finaldeltaW}) to obtain
the expression
for $\Delta F(T)$
in (\ref{DFT}).


\begin{references}


\bibitem{kolb} E. W. Kolb and M. S. Turner, {\it The Early Universe}
(Addison-Wesley,
Redwood, CA, 1990).

\bibitem{qcd}  E. Witten, {\sl Phys. Rev.} {\bf D30}, 272 (1984);
        J. H. Appelgate and C. J. Hogan, {\sl Phys. Rev.} {\bf D31},
        3037 (1985).

\bibitem{weak} For a recent review see A. Cohen, D. Kaplan, and
A. Nelson, Boston University preprint No.
BUHEP-93-4, in press {\it Ann. Rev. Part. Nucl. Phys.}.

\bibitem{gunton} J. D. Gunton, M. San Miguel and P. S. Sahni, in
{\it Phase Transitions and
Critical Phenomena}, {\bf Vol. 8}, Ed. C. Domb and J. L. Lebowitz (Academic
Press, London, 1983);

\bibitem{cahn} J. W. Cahn and J. E. Hilliard,
             {\sl J. Chem. Phys.} {\bf 31}, 688 (1959).


\bibitem{langer67} J. S. Langer, {\it Ann. Phys. (NY)} {\bf 41}, 108 (1967);
{\it ibid.} {\bf 54}, 258 (1969);

\bibitem{alford} M. G. Alford and M. Gleiser, Dartmouth College preprint
No. DART-HEP/93-03; M. G. Alford, H. Feldman, and M. Gleiser,
{\it Phys. Rev.}
{\bf D47} (RC), 2168 (1993).

\bibitem{volosh} M. B. Voloshin, I. Yu. Kobzarev, and L. B. Okun',
        {\sl Yad. Fiz.} {\bf 20}, 1229 (1974)
        [Sov. J. Nucl. Phys. {\bf 20}, 644 (1975) ].


\bibitem{coleman} S. Coleman, {\it Phys. Rev.} {\bf D15}, 2929 (1977);
C. Callan and S. Coleman, {\it Phys. Rev.} {\bf D16}, 1762 (1977);

\bibitem{linde} A. D. Linde,  {\it Phys. Lett.} {\bf 70B}, 306 (1977);
{\it Nucl. Phys.} {\bf B216}, 421 (1983);
[Erratum: {\bf
B223}, 544 (1983)];

\bibitem{affleck} I. Affleck, {\it Phys. Rev. Lett.} {\bf 46}, 388 (1981).


\bibitem{csernai} L. P. Csernai and J. I. Kapusta, {\it Phys. Rev.} {\bf D46},
1379 (1992);

\bibitem{buch} W. Buchm${\ddot {\rm u}}$ller, T. Helbig, and D. Walliser,
               DESY preprint, No. 92-151;  W. Buchm${\ddot {\rm u}}$ller,
               Z. Fodor, T. Helbig, and D. Walliser,
               DESY preprint, No. 93-021.


\bibitem{leggett} See, for example,
                 H. Grabert, U. Weiss, and P. Hanggi, {\it Phys. Rev. Lett.}
                 {\bf 52}, 2193 (1984).

\bibitem{igor} E. W. Kolb and I. Tkachev, {\it Phys. Rev.} {\bf D46}, 4235
               (1992).

\bibitem{kapusta} J. I. Kapusta, {\it Finite Temperature Field Theory},
                  (Cambridge University Press, Cambridge, 1989).

\bibitem{gkw}
            M. Gleiser and E. W. Kolb, {\it Phys. Rev. Lett.} {\bf 69},
            1304 (1992); FERMILAB preprint No. Pub-92/222-A; {\it Int. J.
            Mod. Phys.} {\bf C3}, 773 (1992); M. Gleiser, E. W. Kolb, and
            R. Watkins, {\it Nucl. Phys.} {\bf B364}, 411 (1991).



\bibitem{weinb} E. J. Weinberg, Columbia Univ. Preprint
{\bf CU-TP-577} (1992);

\bibitem{cgm} S. Coleman, V. Glaser, and A. Martin, {\it Commun. Math.
Phys.} {\bf 58}, 211 (1978).

\bibitem{ramos}  G. C. Marques and R. O. Ramos, {\it Phys. Rev.} {\bf D45},
4400 (1992);

\bibitem{mclerran} P. Arnold and L. Mclerran, {\it Phys. Rev.} {\bf D36},
581 (1987);

\bibitem{jackiw} L. Dolan and R. Jackiw, {\it Phys. Rev.} {\bf D9}, 3320
(1974).

\bibitem{gil} C. A. Carvalho, G. C. Marques, A. J. Silva and I. Ventura,
{\it Nucl. Phys.} {\bf B265}, 45 (1986);
C. A. Carvalho, D. Bazeia, O. J. P. Eboli, G. C. Marques, A. J. Silva
and I. Ventura {\it Phys.
Rev.} {\bf D31}, 1411 (1985);

\bibitem{gleiser} M. Gleiser, {\it Phys. Rev.} {\bf D42}, 3350 (1990).
\label{gleiser}

\end{references}
\end{document}